\documentclass[sigconf,authorversion=true]{acmart}

\usepackage{booktabs} 

\usepackage{ifpdf}
\ifpdf
	\graphicspath{{./figs/}}
	\DeclareGraphicsExtensions{.pdf,.jpeg,.png}
\else
	\DeclareGraphicsExtensions{.eps}
\fi

\usepackage{subfig}

\usepackage{algorithm}
\usepackage{algpseudocode}
\algdef{SE}[DOWHILE]{Do}{doWhile}{\algorithmicdo}[1]{\algorithmicwhile\ #1}%

\usepackage{rotating}
\usepackage{multirow}

\usepackage{tikz}

\usepackage[colorinlistoftodos]{todonotes}

\usepackage{url}

\setcopyright{rightsretained}

\acmDOI{10.475/123_4}

\acmISBN{123-4567-24-567/08/06}

\acmConference[]{}{}{} 
\acmYear{2017}
\copyrightyear{2017}

\begin{document}
\sloppy

\title{Flat ORAM: A Simplified {Write-Only} Oblivious RAM Construction for Secure Processors}


\author{Syed Kamran Haider}
\affiliation{\institution{University of Connecticut}}
\email{syed.haider@uconn.edu}

\author{Marten van Dijk}
\affiliation{\institution{University of Connecticut}}
\email{marten.van\_dijk@uconn.edu}


\begin{abstract}

Oblivious RAM (ORAM) is a cryptographic primitive which obfuscates the access patterns to a storage thereby preventing privacy leakage. 
So far in the current literature, only `fully functional' ORAMs are widely studied which can protect, at a cost of considerable performance penalty, against the strong adversaries who can monitor all read and write operations.
However, recent research has shown that information can still be leaked even if only the write access pattern (not reads) is visible to the adversary.
For such weaker adversaries, a fully functional ORAM turns out to be an overkill causing unnecessary overheads. 
Instead, a simple `write-only' ORAM is sufficient, and, more interestingly, is preferred as it can offer far more performance and energy efficiency than a fully functional ORAM.

In this work, we present Flat ORAM: an efficient write-only ORAM scheme which outperforms the closest existing write-only ORAM called HIVE.
HIVE suffers from performance bottlenecks while managing the memory occupancy information vital for correctness of the protocol.
Flat ORAM resolves these bottlenecks by introducing a simple idea of Occupancy Map (OccMap) which efficiently manages the memory occupancy information resulting in far better performance.
Our simulation results show that, on average, Flat ORAM only incurs a moderate slowdown of $3\times$ over the insecure DRAM for memory intensive benchmarks among Splash2 and $1.6\times$ for SPEC06.
Compared to HIVE, Flat ORAM offers $50\%$ performance gain on average and up to $80\%$ energy savings.

\end{abstract}

%
%



\maketitle


\section{Introduction} \label{sec:intro}

User's data privacy concerns in computation outsourcing to cloud servers have gained serious attention over the past couple of decades.
To address this challenge, various \textit{trusted-hardware} based secure processor architectures have been proposed including TPM~\cite{arbaugh97secure,VirtualCountersSTC06, tcg-spec04}, Intel's TPM+TXT~\cite{grawrock-book}, eXecute Only Memory (XOM)~\cite{xom-modelcheck,xom-os,xom-2000}, Aegis~\cite{aegis_processor, aegis_impl}, Ascend~\cite{ascend-stc12}, Phantom~\cite{phantom} and Intel's SGX~\cite{intelSGX}.
A trusted-hardware platform receives user's encrypted data, which is decrypted and computed upon inside the trusted boundary, and finally the encrypted results of the computation are sent to the user.

Although all the data stored outside the secure processor's trusted boundary, e.g. in the main memory, can be encrypted, information can still be leaked to an adversary through the access patterns to the stored data~\cite{ZHUANG04}.
In order to prevent such leakage, Oblivious RAM (ORAM) is a well established technique, first proposed by Goldreich and Ostrovsky~\cite{GO96}.
ORAM obfuscates the memory access pattern by introducing several randomized `redundant' accesses, thereby preventing privacy leakage at a cost of significant performance penalty.
Intense research over the last few decades has resulted in more and more efficient and secure ORAM schemes \cite{oramDMP, oram11, GMOT11, GMOT12, oram12a, OsORAM, oram97, SCSL11, SSS12, PathORAM, oram12c, oram-asplos15, forkpath, yu2015proram}.

A key observation regarding the adversarial model assumed by the current renowned ORAM techniques is that the adversary is capable of learning fine-grained information of \textit{all} the accesses made to the memory.
This includes the information about which location is accessed, the type of operations (read/write), and the time of the access.
It is an extremely strong adversarial model which, in most cases, requires direct physical access to the memory address bus in order to monitor both read \textit{and} write accesses, e.g. the case of a \textit{curious} cloud server.

On the other hand, for purely remote adversaries (where the cloud server itself is trusted), direct physical access to the memory address bus is not possible thereby preventing them from directly monitoring read/write access patterns.
Such remote adversaries, although weaker than adversaries having physical access, can still ``learn'' the application's write access patterns.
Interestingly, privacy leakage is still possible even if the adversary is able to infer just the write access patterns of an application.

John \emph{et al.} demonstrated such an attack \cite{dma_attack_tara} on the famous Montgomery's ladder technique~\cite{montgomery} commonly used for modular exponentiation in public key cryptography.
In this attack, a $512$-bit secret key is correctly inferred in just $3.5$ minutes by only monitoring the application's write access pattern via a compromised Direct Memory Access (DMA\footnote{DMA is a standard performance feature which grants full access of the main memory to certain peripheral buses, e.g. FireWire, Thunderbolt etc.})~\cite{blass2012tresor, bock2009firewire, maartmann2011inception, boileau2006hit, panholzer2008physical} device on the system.
The adversary collects the \textit{snapshots} of the application's memory via the compromised DMA.
Clearly, any two memory snapshots only differ in the locations where the data has been modified in the latter snapshot.
In other words, comparing the snapshots not only reveals the fact that write accesses (if any) have been made to the memory, but it also reveals the exact locations of the accesses which leads to a precise access pattern of memory writes resulting in privacy leakage.

Recent work \cite{DMAmalware} demonstrated that DMA attacks can also be launched \textit{remotely} by injecting malware to the dedicated hardware devices, such as graphic processors and network interface cards, attached to the host platform.
This allows even a remote adversary to learn the application's write access pattern. 
Intel's TXT has also been a victim of DMA based attacks where a malicious OS directed a network card to access data in the protected VM~\cite{TXTattack1, TXTattack2}.

One approach to prevent such attacks, as adapted by TXT, could be to block certain DMA accesses through modifications in DRAM controller.
However, this requires the DRAM controller to be included in the trusted computing base (TCB) of the system which is undesirable.
Nevertheless, there could potentially be many scenarios other than DMA based attacks where write access patterns can be learned by the adversary.

Current so-called \textit{fully functional} ORAM schemes, which obfuscate both read \textit{and} write access patterns, also offer a solution to such weaker adversaries.
However, the added protection (obfuscation of reads) offered by fully functional ORAMs is unnecessary and is practically an overkill in this scenario which results in significant performance penalty.
Path ORAM~\cite{PathORAM}, the most efficient and practical fully functional ORAM system for secure processors so far, still incurs about $2-10\times$ performance degradation~\cite{ascend-stc12, oram-isca13} over an insecure DRAM system.
A far more efficient solution to this problem is a \textit{write-only} ORAM scheme, which only obfuscates the write accesses made to the memory.
Since read accesses leave no trace in the memory snapshot
and hence do not need to be obfuscated in this model, a write-only ORAM can offer significant performance advantage over a fully functional ORAM.

A recent work, HIVE~\cite{hive-ccs14}, has proposed a write-only ORAM scheme for implementing hidden volumes in hard disk drives.
The key idea is similar to Path ORAM, i.e., each data block is written to a new random location, along with some dummy blocks, every time it is accessed.
However, a fundamental challenge that arises in this scheme is to avoid \textit{collisions}.
I.e., to determine whether a randomly chosen physical location contains \textit{real} or \textit{dummy} data, in order to avoid overwriting the existing useful data block.
HIVE proposes a complex \textit{inverse position map} approach for collision avoidance.
Essentially, it maintains a forward position mapping (logical to physical blocks) and a backward/inverse position mapping (physical to logical blocks) for the whole memory. 
Before each write, both forward and backward mappings are looked up to determine whether the corresponding physical location is vacant or occupied.
This approach, however, turns out to be a storage and performance bottleneck because of the size of inverse position map and dual lookups of the position mappings.

A simplistic and obvious solution to this long-standing problem, from a computer architect's perspective, would be to use a bit-mask to mark each physical block as vacant or occupied.
Based on this intuition, we propose a simplified write-only ORAM scheme called Flat ORAM\footnote{Flat ORAM replaces the binary tree structure of Path ORAM with a \textit{flat} array of data blocks; hence termed as `Flat' ORAM.}.
At the core of the algorithm, Flat ORAM introduces a new data structure called Occupancy Map (OccMap): a bit-mask which offers an efficient collision avoidance mechanism.
The OccMap records the availability information (occupied/vacant) of every physical location in a highly compressed form (i.e., just $1$-bit per cache line).
For typical parameter settings, OccMap is about $25\times$ compact compared to HIVE's inverse position map structure.
This dense structure allows the OccMap blocks to exploit higher locality, resulting in considerable performance gain.
While naively storing the OccMap breaks the ORAM's security, we present how to securely store and manage this structure in a real system.
\noindent In particular, the paper makes the following contributions:
\begin{enumerate}
	\item We are the first ones to implement an existing write-only ORAM scheme, HIVE, 
	in the context of secure processors with all state-of-the-art ORAM optimizations, 
	and analyze its performance bottlenecks.
 
	\item A simple write-only ORAM, named Flat ORAM, having an efficient collision avoidance approach is proposed. 
	The micro-architecture of the scheme is discussed in detail and the design space is comprehensively explored.
	It has also been shown to seamlessly adopt various performance optimizations of its predecessor: Path ORAM. 

	\item Our simulation results show that, on average, Flat ORAM offers ${50\%}$ performance gain (up to $75\%$ for DBMS) 
	over HIVE, and only incurs slowdowns of $3\times$ and $1.6\times$ over the 
	insecure DRAM for memory bound Splash2 and SPEC06 benchmarks respectively.
	 
\end{enumerate}

The rest of the paper is organized as follows:
Section~\ref{sec:adversarial_model} describes our adversarial model in detail along with a practical example from the current literature.
Section~\ref{sec:background} provides the necessary background of fully functional ORAMs and write-only ORAMs.
The proposed Flat ORAM scheme along with its security analysis is presented in Section~\ref{sec:flat-algo}, and the detailed construction of its occupancy map structure is shown in Section~\ref{sec:detailed-arch}.
A few additional optimizations from literature implemented in Flat ORAM are discussed in Section~\ref{sec:existing_tricks}.
Section~\ref{sec:eval} evaluates Flat ORAM's performance, and we conclude the paper in Section~\ref{sec:conclusion}.

\section{Adversarial Model} \label{sec:adversarial_model}
We assume a relaxed form of the adversarial model considered in several prior works related to fully functional oblivious RAMs in secure processor settings~\cite{oram-isca13, yu2015proram, oram-asplos15}.

In our model, a user's computation job is outsourced to a cloud, where a trusted processor performs the computation on user's private data.
The user's private data is stored (in encrypted form) in the untrusted memory external to the trusted processor, i.e. DRAM.
In order to compute on user's private data, the trusted processor interacts with DRAM.
The cloud service itself is not considered as an adversary, i.e. it does not try to infer any information from the memory access patterns of the user's program.
However, since the cloud serves several users at the same time, sharing of critical resources such as DRAM among various applications from different users is inevitable.
Among these users being served by the cloud service, we assume a malicious user who is able to monitor remotely (and potentially learn the secret information from) the data write sequences of other users' applications to the DRAM, e.g., by taking frequent snapshots of the victim application's memory via a compromised DMA.
Moreover, he may also tamper with the DRAM contents or play replay attacks in order to manipulate other users' applications and/or learn more about their secret data.

To protect the system from such an adversary, we add to the processor chip a \textit{Write-Only ORAM controller}: an additional trusted hardware module.
Now all the off-chip traffic goes to DRAM through the ORAM controller.
In order to formally define the security properties satisfied by our ORAM controller, we adapt the write-only ORAM \textit{privacy} definition from \cite{haider2017revisiting} as follows:


\begin{definition}
\textbf{(Write-Only ORAM Privacy)}\label{def:privacy}
For every two logical access sequences $A_1$ and $A_2$ of infinite length, their corresponding (infinite length) probabilistic access sequences $\mathsf{ORAM}(A_1)$ and $\mathsf{ORAM}(A_2)$ are identically distributed in the following sense: For all positive integers $n$, if we truncate $\mathsf{ORAM}(A_1)$ and $\mathsf{ORAM}(A_2)$ to their first $n$ accesses, then the truncations $[\mathsf{ORAM}(A_1)]_n$ and $[\mathsf{ORAM}(A_2)]_n$ are identically distributed.
\end{definition}

In other words, memory snapshots only reveal to the adversary the \emph{timing} of write accesses made to the memory (i.e. leakage over \emph{ORAM Timing Channel})
instead of their precise access pattern, whereas no trace of any \textit{read} accesses made to the memory is revealed to the adversary.
An important aspect of Definition~\ref{def:privacy} to note is that it completely isolates the problem of leakage over \emph{ORAM Termination Channel} from ORAM's originally targeted problem (which is also targeted in this paper) i.e., preventing leakage over memory address channel.
Notice that the original definition of ORAM~\cite{GO96} does not protect against timing attacks, i.e. it does not obfuscate \emph{when} an access is made to the memory (ORAM Timing Channel) or how long it takes for the application to finish (ORAM Termination Channel).
The write-only ORAM security definition followed by HIVE~\cite{hive-ccs14} also allows leakage over ORAM termination channel, as two memory access sequences generated by the ORAM for two same-length logical access sequences can have different lengths~\cite{haider2017revisiting}. 
Therefore, in order to define precise security guarantees offered by our ORAM, we follow Definition~\ref{def:privacy}. 

Periodic ORAMs~\cite{ascend-stc12} deterministically make ORAM accesses always at regular predefined (publicly known) intervals, thereby preventing leakage over ORAM timing channel and shifting it to the ORAM termination channel.
We present a periodic variant of our write-only ORAM to protect leakage over ORAM timing channel.
Following the prior works~\cite{phantom}~\cite{oram-asplos15}, (a) we assume that the timing of individual DRAM accesses made \textit{during} an ORAM access does not reveal any information; (b) we do not protect the leakage over ORAM termination channel (i.e. total number of ORAM or DRAM accesses).
The problem of leakage over ORAM termination channel has been addressed in the existing literature~\cite{leakage-hpca14} where only $\log_2(n)$ bits are leaked for a total of $n$ accesses.
A similar approach can easily be applied to the current scheme.


In order to detect malicious tampering of the memory by the adversary, we follow the standard definition of data \textit{integrity} and \textit{freshness}~\cite{oram-asplos15}:
\begin{definition}\label{def:integrity}
\textbf{(Write-Only ORAM Integrity)}
From the processor's perspective, the ORAM behaves like a valid memory with overwhelming probability, and detects any violation to data authenticity and/or freshness.
\end{definition}

\subsection{Practicality of the Adversarial Model}

\begin{algorithm}[!t]
\caption{Montgomery Ladder} \label{algo:ladder}
\begin{flushleft}
\textbf{Inputs:} $g$, $k=(k_{t-1}, \cdots , k_0)_2$ \;\;\; \textbf{Output:} $y=g^k$ \\
\textbf{Start:}
\end{flushleft}
\begin{algorithmic}
	\State $R_0 \gets 1$; $R_1 \gets g$ 
	\For{$j=t-1, 0$}
		\If{$k_j = 0$} ~~ $R_1 \gets R_0 R_1$; ~ $R_0 \gets (R_0)^2$
		\Else ~~ $R_0 \gets R_0 R_1$; ~ $R_1 \gets (R_1)^2$
		\EndIf
	\EndFor
\end{algorithmic}
\begin{flushleft}
\textbf{return} $R_0$
\end{flushleft}
\end{algorithm}

Modular exponentiation algorithms, such as RSA algorithm, are widely used in public-key cryptography. 
In general, these algorithms perform computations of the form $y = g^k\mod n$, where the attacker's goal is to find the secret $k$.
Algorithm~\ref{algo:ladder} shows the Montgomery Ladder scheme~\cite{montgomery} which performs exponentiation ($g^k$) through simple square-and-multiply operations.
For a given input $g$ and a secret key $k$, the algorithm performs multiplication and squaring operations on two local variables $R_0$ and $R_1$ for each bit of $k$ starting from the most significant bit down to the least significant bit.
This algorithm prevents leakage over power side-channel since, regardless of the value of bit $k_j$, the same number of operations are performed in the same order, hence producing the same power footprint for $k_j = 0$ and $k_j=1$.

Notice, however, that the specific order in which $R_0$ and $R_1$ are updated in time depends upon the value of $k_j$.
E.g., for $k_j = 0$, $R_1$ is written first and then $R_0$ is updated; whereas for $k_j = 1$ the updates are done in the reverse order.
This sequence of write access to $R_0$ and $R_1$ reveals to the adversary the exact bit values of the secret key $k$.
A recent work \cite{dma_attack_tara} demonstrated such an attack where frequent memory snapshots of victim application's data (particularly $R_0$ and $R_1$) from the physical memory are taken via a compromised DMA.
These snapshots are then correlated in time to determine the sequence of write access to $R_0$, $R_1$, which in turn reveals the secret key.
The reported time taken by the attack is $3.5$ minutes.

One might argue that under write-back caches, the updates to application's data will only be visible in DRAM once the data is evicted from the LLC.
This will definitely introduce noise to the precise write-access sequence discussed earlier, hence making the attacker's job difficult.
However, he can still collect several `noisy' sequences of memory snapshots and then run correlation analysis on them to find the secret key $k$.
Furthermore, if the adversary is also a user of the same computer system, he can flush the system caches frequently to reduce the noise in write-access sequence even further.

\section{Background of Oblivious RAMs} \label{sec:background}



A fully functional Oblivious RAM \cite{GO96}, or more commonly known as ORAM, is a primitive that obfuscates the user's (i.e. Processor's) access patterns to a storage (i.e. DRAM) such that by monitoring the memory access patterns, an adversary is not able to learn anything about the data being accessed.
The ORAM interface transforms the user's access sequence of program addresses 
into a sequence of ORAM accesses to random looking physical addresses.
Since the physical locations being accessed are revealed to the adversary, the ORAM interface guarantees that the physical access pattern 
is independent of the logical access pattern 
hence user's potentially data dependent access patterns are not revealed.
Furthermore, the data stored in ORAMs should be encrypted using probabilistic encryption to conceal the data content as well as the fact whether or not the content has been updated.

\subsection{Path ORAM} \label{sec:basic-pathoram}
Path ORAM~\cite{PathORAM} is currently the most efficient and well studied ORAM implementation for secure processors.
It has two main hardware components: the \emph{binary tree storage} and the \emph{ORAM controller}. 
{Binary tree} stores the data content of the ORAM and is implemented on DRAM. 
Each node in the tree can hold up to $Z$ useful data blocks, and any empty slots are filled with dummy blocks.
All blocks, real or dummy, are probabilistically encrypted and cannot be distinguished.
The path from the root node to the leaf $s$ is defined as path $s$.
{ORAM controller} is a piece of trusted hardware that controls the tree structure.
Besides necessary logic circuits, the ORAM controller contains two main structures, a \emph{position map} 
and a \emph{stash}.
The \emph{position map} is a lookup table that associates the program address $a$ of a data block with a path in 
the ORAM tree (path $s$).
The \emph{stash} is a buffer that stores up to a small number of data blocks at a time.

Each data block $a$ in Path ORAM is mapped (randomly) to some path $s$ via the position map, i.e. at any time, the data block $a$ must be stored either on path $s$, or in the stash. 
Path ORAM follows the following steps when a request on block $a$ is issued by the processor:
(1) The path (leaf) number $s$ of the logical address $a$ is looked up in the position map. 
(2) All the blocks on path $s$ are read and decrypted, and all real blocks added to the stash.
(3) Block $a$ is returned to the processor.
(4) The position map of $a$ is updated to a new random leaf $s'$. 
(5) As many blocks from stash as possible are encrypted and written on path $s$, where empty spots are filled with dummy blocks.
Step (4) guarantees that when block $a$ is accessed later, a random path will be accessed which is independent of any previously accessed paths (\textit{unlinkability}). 
As a result, each ORAM access is random and unlinkable regardless of the request pattern.

\label{sec:limitation}
Path ORAM incurs significant energy and performance penalties compared to insecure DRAM.  
Under typical settings for secure processors (gigabytes of memory and 64- to 128-byte blocks), Path ORAM has a 20-30 level binary tree where each node typically stores 3 or 4 data blocks~\cite{SSS12, oram-isca13}.  
This means that each ORAM access reads \textit{and} writes 60-120 blocks, in contrast to a single read \textit{or} write operation in an insecure storage system.



\subsection{Write-Only ORAMs}\label{sec:wo-orams}

In contrast to fully functional ORAMs, a write-only ORAM only obfuscates the patterns of \textit{write} accesses made to a storage.
Write-only ORAM is preferred for performance reasons over fully functional ORAMs in situations where the adversary is not able to monitor the \textit{read} access patterns.

There has been very limited research work done so far to explore write-only ORAMs, and to best of our knowledge, write-only ORAMs for secure processors have not been explored at all.
Li and Datta~\cite{2013writeonly} present a write-only ORAM scheme to be used with Private Information Retrieval (PIR) in order to preserve the privacy of data outsourced to a data center.
Although this scheme achieves an amortized write cost of $O(B\log N)$, it incurs a read cost of $O(B.N)$ for a storage of $N$ blocks each of size $B$.
For efficient reads, it requires the client side storage (i.e. the on-chip position map) to be polynomial in $N$.
In a secure processor setting, DRAM reads are usually the major performance bottleneck, and introducing a complexity polynomial in $N$ on this critical path is highly unwanted.

A recent work, HIVE~\cite{hive-ccs14}, has proposed a write-only ORAM scheme for hidden volumes in hard disk drives.
Although HIVE write-only ORAM presented-as-is \cite{hive-ccs14} targets a totally different application, we believe that its parameter settings can be tweaked to be used in the secure processor setting.
This paper is the first one to implement HIVE in the secure processor context, and we consider this implementation as the baseline write-only ORAM to be compared with our proposed Flat ORAM.

Roche \emph{et al.}~\cite{roche2017deterministic} have also proposed an efficient write-only ORAM scheme for hard disk storages. 
This scheme along with its complex optimizations has been implemented in software which is completely feasible at the DRAM-Disk boundary. 
However, in this work, we target the Processor-DRAM boundary for our proposed ORAM which needs to be implemented in hardware, and hence our focus is only towards simplified algorithms and optimizations which can easily be synthesized in hardware without substantial area overhead.

\section{Flat ORAM Scheme}\label{sec:flat-algo}

In this section, we first present the core algorithm of Flat ORAM, then we discuss its architectural details and various optimizations for a practical implementation. 

\subsection{Fundamental Data Structures}

\paragraph{Position Map (PosMap):} 
It is a standard Path ORAM structure that maintains randomized mappings of logical blocks to physical locations.
However there is one subtle difference between PosMap of Path ORAM and Flat ORAM.
In Path ORAM, PosMap stores a path number for each logical block and the block can reside \textit{anywhere} on that path.
In contrast, PosMap in Flat ORAM stores the exact physical address where a logical block is stored.

\paragraph{Occupancy Map (OccMap):} 
OccMap is a newly introduced structure in Flat ORAM.
It is essentially a large bit-mask where each bit corresponds to a physical location (i.e., a cache line).
The binary value of each bit represents whether the corresponding physical block contains real or outdated/dummy data.
OccMap is of crucial importance to avoid \textit{data collisions}, and hence for the correctness of Flat ORAM.
A collision happens when a physical location, which is randomly chosen to store a logical block, already contains useful data which cannot be overwritten.
Managing the OccMap securely and efficiently is a major challenge which we address in section~\ref{sec:detailed-arch} in detail.

\paragraph{Stash:}\label{subsec:stash}
Stash, also adapted from Path ORAM, is a small buffer in the trusted memory to temporarily hold data blocks evicted from the processor's last level cache (LLC).
A slight but crucial modification, however, is that Flat ORAM only buffers \textit{dirty}\footnote{Blocks with modified data.} data blocks in the stash; while \textit{clean} blocks evicted from the LLC are simply ignored since a valid copy of these blocks already exists in the main memory.
This modification is significantly beneficial for performance.

\subsection{Basic Algorithm}\label{sec:basic-algo}

\begin{algorithm}[!t]
\caption{Flat ORAM Initialization.} \label{algo:Initial}
\begin{algorithmic}[1]
\Procedure{Initialize}{ $N, B, P$ }
	\State $\textsf{PosMap} := \{\bot \}^N $	\Comment{Empty Position Map.}
	\State $\textsf{OccMap} := \{0\}^P $	\Comment{Empty Occupancy Map.}
	\For{$j \in \{1, \cdots, N\}$}
		\Loop
			\State $r \gets$ \Call{UniformRand}{$1, \cdots, P$}
			\If{$\textsf{OccMap}[r] == 0$}			\Comment{If vacant}
				\State $\textsf{OccMap}[r] := 1$		\Comment{Mark Occupied.}
				\State $\textsf{PosMap}[j] := r$			\Comment{Record position.}
				\State \textbf{break}
			\EndIf
		\EndLoop
	\EndFor
\EndProcedure
\end{algorithmic}
\end{algorithm}

Let $N$ be the total number of logical data blocks that we want to securely store in our ORAM, which is implemented on top of a DRAM; and let each data block be of size $B$ bytes.
Let $P$ be the number of physical blocks that our DRAM can physically store, i.e. the DRAM capacity (where $P\ge N$).

\textbf{Initial Setup:}
Algorithm~\ref{algo:Initial} shows the setup phase of our scheme.
Two null-initialized arrays $\textsf{PosMap}$ and $\textsf{OccMap}$, corresponding to position and occupancy map of size $N$ and $P$ entries respectively, are allocated.
For now, we assume that both $\textsf{PosMap}$ and $\textsf{OccMap}$ reside on-chip in the trusted memory to which the adversary has no access.
However, since these arrays can be quite large and the trusted memory is quite constrained, we later on show how this problem is solved.
Initially, since all physical blocks are empty, each OccMap entry is set to $0$.
Now, each logical block is mapped to a uniformly random physical block, i.e. PosMap is initialized, while avoiding any collisions using OccMap.
The OccMap is updated along the way in order to mark those physical locations which have been assigned to a logical block as `occupied'.
Notice that in order to minimize the probability of collision, $P$ should be sufficiently larger than $N$, e.g. $P \approx 2N$ gives a $50 \%$ collision probability.

\begin{algorithm}[!t]
\caption{Basic Flat ORAM considering all PosMap and OccMap is on-chip.
Following procedures show reading, writing and eviction of a logical block $a$ from the ORAM.} \label{algo:basic-oram}
\begin{algorithmic}[1]
\Procedure{ORAMRead}{$a$}
	\State $s := \textsf{PosMap}[a] $		\Comment{Lookup position}
	\State $data := \mathsf{Dec_{K}}(\Call{DRAMRead}{$s$}) $
	\State \textbf{return} $(s, data)$	\Comment{Position is also returned.}
\EndProcedure
\end{algorithmic}

\vspace{4pt}
\begin{algorithmic}[1]
\Procedure{ORAMWrite}{$a, s_{\mathtt{old}}, data$}
	\State $Stash := Stash \cup \{ (a, s_{\mathtt{old}}, data) \} $ \Comment{Add to Stash}
	\State \textbf{return}
\EndProcedure
\end{algorithmic}

\vspace{4pt}
\begin{algorithmic}[1]
\Procedure{EvictStash}{}
	\State $(a, s_{\mathtt{old}}, data) \gets Stash$ \Comment{Read from Stash}
	\Loop
		\State $s_{\mathtt{new}} \gets$ \Call{UniformRand}{$1, \cdots, P$}
		\If{$\textsf{OccMap}[s_{\mathtt{new}}] == 0$}			\Comment{If vacant}
			\State $\textsf{OccMap}[s_{\mathtt{new}}] := 1$	\Comment{Mark as Occupied.}
			\State $\textsf{OccMap}[s_{\mathtt{old}}] := 0$		\Comment{Vacate old block.}
			\State $\textsf{PosMap}[a] := s_{\mathtt{new}}$		\Comment{Record position.}
			\State \Call{DRAMWrite}{$s_{\mathtt{new}}, \mathsf{Enc_{K}}(data)$}
			\State $Stash := Stash \setminus \{ (a, s_{\mathtt{old}}, data) \}$		
			\State \textbf{break}
		\Else																	\Comment{If occupied.}
			\State $data' := \mathsf{Dec_{K}}(\Call{DRAMRead}{s_{\mathtt{new}}}) $
			\State \Call{DRAMWrite}{$s_{\mathtt{new}}, \mathsf{Enc_{K}}(data')$}
		\EndIf
	\EndLoop
\EndProcedure
\end{algorithmic}
\end{algorithm}

\textbf{Reads:}
The procedures to read/write a data block corresponding to the virtual address $a$ from/to the ORAM are shown in Algorithm~\ref{algo:basic-oram}.
A read operation is straightforward as it does not need to be obfuscated.
The PosMap entry for the logical block $a$ is looked up, and the encrypted data is read through normal DRAM read.
The data is decrypted and returned to the LLC along with its current physical position $s$.
The location $s$ is stored in the tag array of the LLC, and proves to be useful upon eviction of the data from the LLC.

\textbf{Writes:}
Since write operations should be non-blocking in a secure processor setting, the ORAM writes are performed in two steps.
Whenever a data block is evicted from the LLC, it is first simply added to the stash queue, without incurring any latency.
While the processor moves on to computing on other data in its registers/caches, the ORAM controller then works in the background to evict the block from stash to the DRAM.
A block $a$ to be written is picked from the stash, and a new uniformly random physical position $s_{\mathtt{new}}$ is chosen for this block. 
The OccMap is looked up to determine whether the location $s_{\mathtt{new}}$ is vacant.
If so, the write operation proceeds by simply recording the new position $s_{\mathtt{new}}$ for block $a$ in PosMap, updating the OccMap entries for $s_{\mathtt{new}}$ and $s_{\mathtt{old}}$ accordingly, and finally writing encrypted data at location $s_{\mathtt{new}}$.
Otherwise if the location $s_{\mathtt{new}}$ is already occupied by some useful data block, the probability of which is $N/P$, the existing data block is read, decrypted, re-encrypted\footnote{We assume that the encryption/decryption algorithms $\mathsf{Enc_{K}}$/$\mathsf{Dec_{K}}$ implement probabilistic encryption, e.g. AES counter mode, as done in prior works \cite{oram-isca13, oram-asplos15}.} under probabilistic encryption and written back.
A new random position is then chosen for the block $a$ and the above mentioned process is repeated until a vacant location is found to evict the block.
Notice that storing $s_{\mathtt{old}}$ along with the data upon reads will save extra ORAM accesses to lookup $s_{\mathtt{old}}$ from the recursive PosMap 
(cf. Section~\ref{sec:recursive-map}).

\subsection{Avoiding Redundant Memory Accesses} \label{sec:redundant_accesses}
The fact that the adversary cannot see read accesses allows Flat ORAM to avoid almost all the redundancy incurred by a fully functional ORAM (e.g. Path ORAM).
Instead of reading/writing a whole path for each read/write access as done in Path ORAM, Flat ORAM simply reads/writes only the desired block directly given its physical location from the PosMap.
This is fundamentally where the write-only ORAMs (i.e. HIVE~\cite{hive-ccs14}, Flat ORAM) get the performance edge over the fully functional ORAMs.
However, the question arises whether Flat ORAM is still secure after eliminating the redundant accesses.

\subsection{Security} \label{sec:security}
\textbf{Privacy}: 
Consider any two logical write-access sequences $O_0$ and $O_1$ of the same length. 
In \Call{EvictStash}{} procedure (cf. Algorithm~\ref{algo:basic-oram}), a physical block chosen uniformly at random out of $P$ blocks is \textit{always} modified regardless of it being vacant or occupied.
Therefore, the write accesses generated by Flat ORAM while executing \emph{either} of the two logical access sequences $O_0$ and $O_1$ will update memory locations uniformly at random throughout the whole physical memory space.
As a result, an adversary monitoring these updates cannot distinguish between \emph{real} vs. \emph{dummy} blocks, and in turn the two sequences $O_0$ and $O_1$ seem computationally indistinguishable.

Furthermore, notice that in Path ORAM the purpose of accessing a whole path instead of just one block upon each read and write access is to prevent \textit{linkability} between a write and a following read access to the same logical block.
In Flat ORAM's model, however, since the adversary cannot see the read accesses at all, therefore the linkability problem would never arise as long as each logical data block is written to a new random location every time it is evicted (which is guaranteed by Flat ORAM algorithm).
Although HIVE proposes a constant $k$-factor redundancy upon each data write, we argue that it is unnecessary for the desired security as explained above, and can be avoided to gain performance.

Hence, the basic algorithm of Flat ORAM presented above guarantees 
the desired \textit{privacy} property of our write-only ORAM (cf. Definition~\ref{def:privacy}).

\vspace{3pt}
\textbf{Integrity}: Next, we move on to making the basic Flat ORAM practical for a real system.
The main challenge is to get rid of the huge on-chip memory requirements imposed by PosMap and OccMap.
While addressing the PosMap management problem, we discuss an existing efficient memory integrity verification technique from Path ORAM domain called $PMMAC$~\cite{oram-asplos15} (cf. Section \ref{sec:integrity}) which satisfies our \textit{integrity} definition (cf. Definition~\ref{def:integrity}).

\vspace{3pt}
\textbf{Stash Management}: Another critical missing piece is to prevent the unlikely event of stash overflow for a small constant sized stash, as such an event could break the privacy guarantees offered by Flat ORAM.
We completely eliminate the possibility of a stash overflow event by using a proven technique called \textit{Background Eviction}~\cite{oram-isca13}.
We present a detailed discussion about the stash size under Background Eviction technique in Section~\ref{sec:stash_size}.

\subsection{Recursive Position Map \& PLB} \label{sec:recursive-map} \label{sec:plb}
In order to squeeze the on-chip PosMap size, a standard recursive construction of position map~\cite{SCSL11} is used.
In a 2-level recursive position map, for instance, the original PosMap structure is stored in another set of data blocks which we call a \textit{hierarchy} of position map, and the PosMap of the first hierarchy is stored in the trusted memory on-chip (\figurename~\ref{fig:occmap}).
The above trick can be repeated, i.e., adding more hierarchies of position map to further reduce the final position map size at the expense of increased latency. 
Notice that all the position map hierarchies (except for the final position map) are stored in the untrusted DRAM along with the actual data blocks, and can be treated as regular data blocks; this technique is called Unified ORAM~\cite{oram-asplos15}.

Unified ORAM scheme reduces the performance penalty of recursion by caching position map ORAM blocks in a Position map Lookaside Buffer (PLB) to exploit locality (similar to the TLB exploiting locality in page tables).
To hide whether a position map access hits or misses in the cache, Unified ORAM stores both data and position map blocks in the same binary tree.
Further compression of PosMap structure is done by using Compressed Position Map technique discussed in section \ref{sec:comp-posmap}.

\subsection{Background Eviction} \label{sec:background-evict}
Stash (cf. Section~\ref{subsec:stash}) is a small buffer to temporarily hold the \textit{dirty} data blocks evicted from the LLC/PLB.
If at any time, the rate of blocks being added to the stash becomes higher than the rate of evictions from the stash, the blocks may accumulate in stash causing a stash overflow.
Background eviction~\cite{oram-isca13} is a proven and secure technique proposed for Path ORAM to prevent stash overflow.
The key idea of background evictions is to temporarily stop serving the real requests (which increase stash occupancy) and issue background evictions or so-called \textit{dummy accesses} (which decrease stash occupancy) when the stash is full.

We use background eviction technique to eliminate the possibility of a stash overflow event. 
When the stash is full, the ORAM controller suspends the read requests which consequently stops any write-back requests preventing the stash occupancy to increase further.
Then it simply chooses random locations and, if vacant, evicts the blocks from the stash until the stash occupancy is reduced to a safe threshold.
The probability of a successful eviction in each attempt is determined by the DRAM \textit{utilization} parameter, i.e. the ratio of occupied blocks to the total blocks in DRAM. 
In our experiments, we choose a utilization of $\approx 1/2$, therefore each eviction attempt has $\approx 50\%$ probability of success.
Note that background evictions essentially push the problem of stash overflow to the program's termination channel.
Configuring the DRAM utilization to be less than $1$ guarantees the termination, and we demonstrate good performance for a utilization of $1/2$ in Section~\ref{sec:eval}.
Although it is true that background eviction may have different effect on the total runtime for different applications, or even for different inputs for the same application which leak the information about data locality etc.
However, the same argument applies to Path ORAM based systems, and also for other system components as well, such as enabling vs. disabling branch prediction or the L3 cache etc.
Protecting any leakage through the program's termination time is out of scope of this paper (cf. Section \ref{sec:adversarial_model}).


%

\subsection{Periodic ORAM} \label{sec:timing-channel}

As mentioned in our adversarial model, the core definition of ORAM \cite{GO96} do not leakage over ORAM timing or termination channel (cf. Section~\ref{sec:adversarial_model}).
Likewise, the fundamental algorithm of Flat ORAM (Algorithm~\ref{algo:basic-oram}) does not target to prevent these leakages.
Therefore in order to protect the ORAM timing channel, we adapt the Flat ORAM algorithm to issue periodic ORAM accesses, while maintaining its security guarantees.

In the literature, periodic variants of Path ORAM have been presented \cite{ascend-stc12} which simply always issue ORAM requests at regular periodic intervals.
However, under write-only ORAMs, such a straightforward periodic approach would break the security as explained below.
Since in write-only ORAMs, the read requests do not leave a trace, therefore for a logical access sequence of \texttt{(Write, Read, Write)}, the adversary will only see two writes occurring at times $0$ and $2T$ for $T$ being the interval between two ORAM accesses.
The access at time $T$ will be omitted which reveals to the adversary that a read request was made at this time.

To fix this problem, we modify the Flat ORAM algorithm as follows.
Among the periodic access, for every real read request to physical block $s$, another randomly chosen physical block $s'$ is also read. 
Block $s$ is consumed by the processor, whereas block $s'$ is re-encrypted (under probabilistic encryption) and written back to the same location from where it was read.
This would always result in update(s) to the memory after each and every time period $T$.

\textbf{Security:} 
A few things should be noted: 
First, it does not matter whether the location $s'$ contains real or dummy data, because the plain-text data content is never modified but just re-encrypted. Second, writing back $s'$ is indistinguishable from a real write request as this location is chosen uniformly at random. 
Third, this write to $s'$ does not reveal any trace of the actual read of $s$ as the two locations are totally independent.

We present our simulation results for periodic Flat ORAM in the evaluation section.

\section{Efficient Collision Avoidance}\label{sec:detailed-arch}

In sections \ref{sec:redundant_accesses} and section \ref{sec:security}, we discuss how Flat ORAM outperforms Path ORAM by avoiding redundant memory accesses.
However, an immediate consequence of this is the problem of \textit{collisions} which now becomes the main performance bottleneck.
\label{sec:collisions}
A collision refers to a scenario when a physical location $s$, which is randomly chosen to write a logical block $a$, already contains useful data which cannot be overwritten.
The overall efficiency of such a write-only ORAM scheme boils down to its collision avoidance mechanism.
In the following subsections, we discuss the occupancy map based collision avoidance mechanism of Flat ORAM in detail and compare it with HIVE's \textit{inverse position map} based collision avoidance scheme.

\vspace{-0.1in}


\subsection{Inverse Position Map Approach}
Since a read access must not leave its trace in the memory in order to avoid the linkability problem, a naive approach of marking the physical location as `vacant' by writing `dummy' data to it upon each read is not possible.
HIVE~\cite{hive-ccs14} proposes an \textit{inverse position map} structure that maps each physical location to a logical address.
Before each write operation to a physical location $s$, a potential collision check is performed which involves two steps.
First the logical address $a$ linked to $s$ is looked up via the inverse position map.
Then the regular position map is looked up to find out the most recent physical location $s'$ linked to the logical address $a$.
If $s=s'$ then since the two mappings are synchronized, it shows that $s$ contains useful data, hence a collision has occurred.
Otherwise, if $s\ne s'$ then this means that the entry for $s$ in the inverse position map is outdated, and block $a$ has now moved to a new location $s'$. 
Therefore, the current location can be overwritten, hence no collision.

HIVE stores the encrypted inverse position map structure in the untrusted storage at a fixed location.
With each physical block being updated, the corresponding inverse position map entry is also updated.
Since write-only ORAMs do not hide the physical block ID of the updated block, therefore revealing the position of the corresponding inverse position map entry does not leak any secret information.

We demonstrate in our evaluations that the large size of inverse position map approach introduces storage as well as performance overheads.
For a system with a block size of $B$ bytes and total $N$ logical blocks, inverse position map requires $\log_2(N)$ additional bits space for each of the $P$ physical blocks.
Crucially, this large size of a single inverse position map entry restricts the total number of entries per block to a small constant, which leads to less locality within a block.
This results in performance degradation.

\subsection{Occupancy Map Approach}
In the simplified OccMap based approach, each of the $P$ physical blocks requires just one additional bit to store the occupancy information (vacant/occupied).
In terms of storage, this gives $\log_2(N)$ times improvement over HIVE.

\subsubsection{Insecurely Managing the Occupancy Map}
The OccMap array bits are first sliced into chunks equal to the ORAM block size ($B$ bytes).
We call these chunks the OccMap Blocks.
Notice that each OccMap block contains occupancy information of $8B$ physical locations (i.e. 8 bits per byte; 1-bit per location).
Now the challenge is to efficiently store these blocks somewhere off-chip.
A naive approach would be to encrypt OccMap blocks under probabilistic encryption, and store them contiguously in a dedicated fixed area in DRAM.
However under Flat ORAM algorithm, this approach would lead to a serious security flaw which is explained below.

If the OccMap blocks are stored contiguously at a fixed location in DRAM, an adversary can easily identify the corresponding OccMap block for a given a physical address; and for a given OccMap block, he can identify the contiguous range of corresponding $8B$ physical locations.
With that in mind, when a data block $a$ (previously read from $s_{\mathtt{old}}$) is evicted from the stash and written to location $s_{\mathtt{new}}$ (cf. Algorithm~\ref{algo:basic-oram}), the old OccMap entry is marked as `vacant' and the new OccMap entry is marked as `occupied'; i.e. two OccMap blocks $O_{\mathtt{old}}$ and $O_{\mathtt{new}}$ are updated.
Furthermore, the $s_{\mathtt{new}}$ location, which falls in the contiguous range covered by one of the two updated OccMap blocks, is also updated with the actual data -- thereby revealing the identity of $O_{\mathtt{new}}$.
This reveals to the adversary that a logical data block was previously read from some location within the small contiguous range covered by $O_{\mathtt{old}}$, and it is now written to location $s_{\mathtt{new}}$ (i.e. coarse grained linkability).
Recording several such instances of read-write access pairs and linking them together in a chain reveals the precise pattern of movement of logical block $a$ across the whole memory.

\subsubsection{Securely Managing the Occupancy Map}
\begin{figure}[!t]
\centering
\includegraphics[width=\columnwidth]{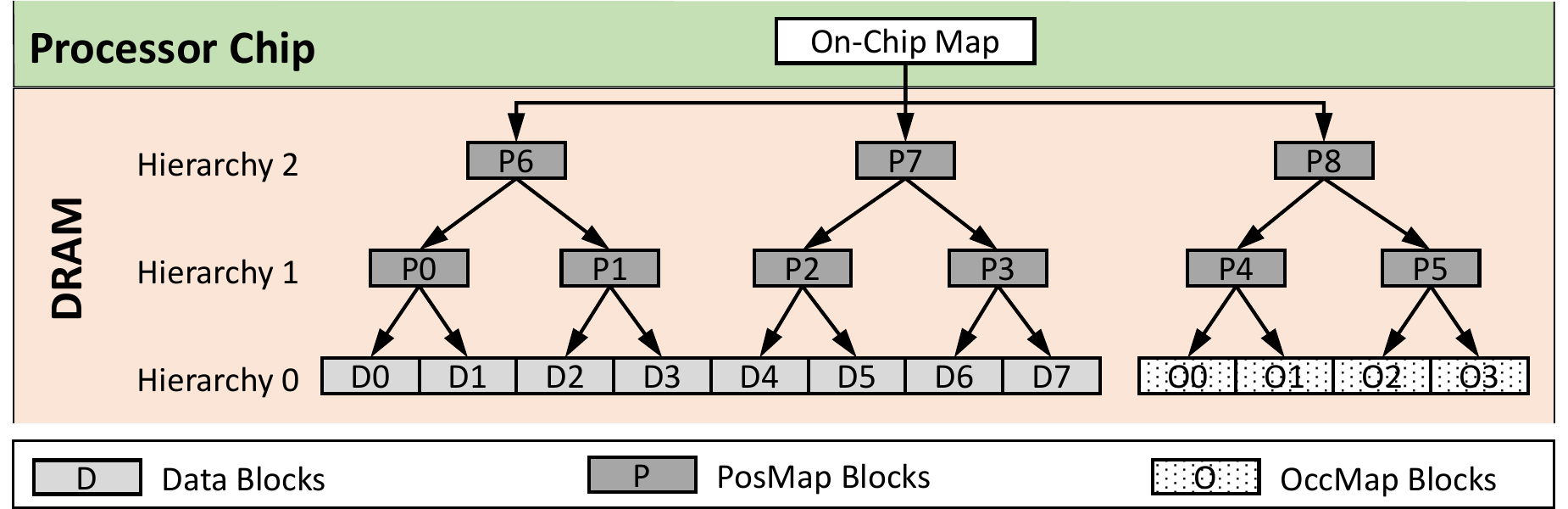}
\caption{Logical view of OccMap organization.}
\label{fig:occmap}
\vspace{-10pt}
\end{figure}

To avoid this problem, we treat the OccMap blocks as regular data blocks, i.e., OccMap blocks are encrypted and also randomly distributed throughout the whole DRAM, and tracked by the regular PosMap.
Figure~\ref{fig:occmap} shows the logical organization of data, PosMap and OccMap blocks.
The OccMap blocks are added as `additional' data blocks at the data hierarchy (Hierarchy 0). 
Then the recursive position map hierarchies are constructed on top of the complete data set (data blocks and OccMap blocks).
Every time an OccMap block is updated, it is mapped to a new random location and hence avoids the linkability problem.
We realize that this approach results in overall more position map blocks, however for practical parameters settings (e.g. $128B$ block size, $8GB$ DRAM, $4GB$ working set), it does not result in an additional PosMap hierarchy.
Therefore the recursive position map lookup latency is unaffected.

\subsection{Performance Related Optimizations}\label{sec:occmap_locality}
Now that we have discussed how to securely store the occupancy map, we move on to discuss some performance related optimizations implemented in Flat ORAM. 

\subsubsection{Locality in OccMap Blocks}
For realistic parameters, e.g. $128$ bytes block size, each OccMap block contains occupancy information of $1024$ physical locations.
This factor is termed as OccMap \textit{scaling factor}.
The dense structure of OccMap blocks offers an opportunity to exploit spatial locality within a block.
In other words, for a large scaling factor it is more likely that two randomly chosen physical locations will be covered by the same OccMap block, as compared to a small scaling factor.
In order to also benefit from such locality, we cache the OccMap blocks as well in PLB along with the PosMap blocks.
For a fair comparison with HIVE in our experiments, we also model the caching of HIVE's inverse position map blocks in PLB.
Our experiments confirm that the OccMap blocks show a higher PLB hit rate as compared to HIVE's inverse position map blocks cached in PLB in the same manner.
The reason is that, for the same parameters, the scaling factor of inverse position map approach is just about $40$ which results in a larger size of the data structure and hence more capacity-misses from the PLB.


\subsubsection{Dirty vs. Clean Evictions from LLC \& PLB}\label{sec:dirty-clean}
An eviction of a block from the LLC where the data content of the block has not been modified is called a \textit{clean} eviction, whereas an eviction where the data has been modified is called \textit{dirty} eviction.
In Path ORAM, since all the \textit{read} operations also need to be obfuscated, therefore following a read operation, when a block gets evicted from the LLC, it must be re-written to a new random location \textit{even if its data is unmodified}, i.e. a clean eviction.
This is crucial for Path ORAM's security as it guarantees that successive reads to the same logical block result in random paths being accessed.
This notion is termed as \textit{read-erase}, which assumes that the data will be erased from the memory once it is read.

In write-only ORAMs, however, since the read access patterns are not revealed therefore the notion of \textit{read-erase} is not necessary.
A data block can be read from the same location as often as needed as long as it's contents are not modified.
We implement this relaxed model in Flat ORAM which greatly improves performance.
Essentially, upon a \textit{clean} eviction from the LLC, the block can simply be discarded since one useful copy of the data is still stored in the DRAM.
Same reasoning applies to the clean evictions from the PLB.
Only the \textit{dirty} evictions are added to the stash to be written back at a random location in the memory.

\subsection{Implications on PLB \& Stash Size} \label{sec:stash_size}

Each dirty eviction requires not only the corresponding data block to be updated but also the two related OccMAP blocks which store the new and old occupancy information. In order to relocate these blocks to new random positions, the `$d$' hierarchies of corresponding PosMAP blocks will need to be updated and this in turn implies updating their related OccMAP blocks, and so on. If not prevented, this avalanche effect will repeatedly fill the stash implying background evictions which stop serving real requests and increase the termination time. For a large enough PLB with respect to a benchmark's {\em locality}, most of the required OccMAP blocks during the benchmark's execution will be in the PLB. This prevents the avalanche effect most of the time (as our evaluation shows) since the OccMAP blocks in PLB can be directly updated.

Even if all necessary OccMAP blocks are in PLB, a dirty eviction still requires the data block with its $d$ PosMAP blocks to be updated. Each of these $d+1$ blocks is successfully evicted from the stash with probability $1/2$, determined by the DRAM utilization, on each attempt. The probability that exactly $m$ attempts are needed to evict all $d+1$ blocks is equal to ${m-1 \choose d}/ 2^{m}$. This probability becomes very small  for $m$ equal to a small constant $c$ times $d\log d$. If the {\em dirty eviction rate} (per DRAM access)  is at most $1/c$, then the stash size will mostly be contained to a small size (e.g., $~100$ blocks for $d=4$) so that additional background eviction which stops serving real requests is not needed. 

Notice that the presented write-only ORAM is not asymptotically efficient: In order to show at most a constant factor increase in termination time with overwhelming probability, a proper argument needs to show a small probability of requiring background eviction which halts normal execution. An argument based on M/D/1 queuing theory or 1-D random walks needs the OccMap to be always within the PLB and this means that the effective stash size as compared to Path ORAM's stash definition includes this PLB which scales linearly with $N$ and is not $O(log N)$.

\section{Adopting More Existing Tricks} \label{sec:existing_tricks}
Here we discuss a few more architectural optimizations from the Path ORAM paradigm which can be flawlessly incorporated and are implemented in Flat ORAM for further improvements and features.

\subsection{Compressed Position Map}\label{sec:comp-posmap}
The recursive position map for a total of $N$ logical data blocks creates $\lceil \log_b(N) \rceil$ hierarchies of position map.
Here $b$ represents the number of positions stored in one PosMap block, and is called \textit{PosMap scale factor}.
A higher value of PosMap scale factor would result in less number of PosMap hierarchies and hence yield better performance.

To achieve this goal, Compressed Position Map~\cite{oram-asplos15} has been proposed, which results in less PosMap hierarchies than uncompressed PosMap.
The basic idea is to store a monotonically increasing counter in the PosMap entry for each logical data block.
This counter along with the block's logical address is used as a `seed' to a keyed pseudo-random function in order to compute a random position for the block.
Every time a block is to be written, its PosMap counter is first incremented so that a new random position is generated by the pseudo-random function for the block.
To compress these counters to a feasible size, \cite{oram-asplos15} presents an optimization using a big \textit{group counter} and several small \textit{individual counters} per PosMap block.
We refer the readers who might be interested in more details to the above citation.

We tweak the compressed PosMap technique for Flat ORAM.
The key modification is that the counter for any block to be evicted is incremented even upon unsuccessful eviction attempts, i.e. even if a collision is detected.
It is important because otherwise the pseudo-random function will generate the same random location over and over which is already occupied, and hence the block will never be evicted.


\subsection{Integrity Verification (PMMAC)}\label{sec:integrity}
Flat ORAM also implements an efficient memory integrity verification technique termed as PosMap MAC (PMMAC)~\cite{oram-asplos15}.
PMMAC leverages the per-block counters of compressed PosMap to perform MAC\footnote{Message Authentication Code (MAC), e.g. a keyed cryptographic hash, is a small piece of information to verify the authenticity of a message/data.} checks on the data upon reads.
Suppose a logical block $a$ has a counter $c$, then upon writes, the ORAM controller computes a MAC $h=\mathsf{MAC}_K(a~||~c~||~data)$ using the secret key $K$ and writes the tuple $(h,data)$ to the DRAM.
Upon reads, the potentially tampered data tuple $(h^*, data^*)$ is read.
The ORAM controller recomputes $h=\mathsf{MAC}_K(a~||~c~||~data^*)$ and checks whether $h=h^*$.
If so, the data integrity is verified.
Also, since the counter is incremented upon every write, the freshness of the data is also verified, i.e. integrity check guarantees that the most recently written data has been read.

\vspace{-0.1in}

\section{Experimental Evaluation} \label{sec:eval}

\subsection{Methodology} \label{sec:method}
\begin{table}
	\renewcommand{\arraystretch}{1.3}
	\caption{ System Configuration. }
	\vspace{-10pt}
	\begin{center}
	{ \footnotesize
		\begin{tabular}{|l|l|}			
			\hline
			\multicolumn{2}{|c|}{Secure Processor Configuration} \\
			\hline
			Core model				& 1~GHz, in order core\\
			Total Cores 				& 4 \\
			L1 I/D Cache				& 32~KB, 4-way \\
			Shared L2 cache 		& 512~KB per tile, 8-way \\
			Cacheline (block) size 	& 128bytes \\
			DRAM bandwidth							& 16 GB/s \\
			Conventional DRAM latency				& 100 cycles \\
			\hline
			\hline
			\multicolumn{2}{|c|}{Default ORAM Configuration} \\
			\hline
			ORAM Capacity								& 8~GB \\
			Working Set Size							& 4~GB \\
			Number of ORAM hierarchies				& 4 \\
			ORAM Block size					 		& 128~Bytes \\
			PLB Size										& 32kB \\
			Stash Size									& 100 Blocks \\
			Compressed PosMap \& Integrity			& Enabled \\
			\hline

		\end{tabular}
	}
	\end{center}
	\vspace{-12pt}
	\label{tab:system}
\end{table}

We use Graphite~\cite{graphite} to model different ORAM schemes in all our experiments.
Graphite simulates a tiled multi-core chip. 
The hardware configurations are listed in Table~\ref{tab:system}.
We assume there is only one memory controller on the chip, and all ORAM accesses are serialized. 
The DRAM in Graphite is simply modeled by a flat latency. 
The 16 GB/s is calculated assuming a 1 GHz chip with 128 pins and pins are the bottleneck of the data transfer. 

We use Splash-2~\cite{Woo:1995}, SPEC06~\cite{spec06} and two OLTP database management system (DBMS)~\cite{yu14} workloads namely YCSB~\cite{cooper10} and TPCC~\cite{tpc-c} to evaluate our Flat ORAM scheme (\textsf{flat\_oram}) against various baselines.
Three baseline designs are used for comparison: the insecure baseline using normal DRAM (\textsf{dram}), the state of the art Path ORAM with dynamic prefetching~\cite{yu2015proram} (\textsf{path\_oram}) and an adaptation of the write-only ORAM scheme from HIVE (\textsf{hive}) in the context of secure processor architectures with several additional optimizations. 
For all ORAM schemes, we enable the PLB, the compressed position map and integrity verification.
The default parameters for ORAM schemes are shown in Table~\ref{tab:system}.  
Unless otherwise stated, all the experiments use these ORAM parameters. 

\subsection{Performance Comparison} \label{sec:eval-splash2}

\begin{figure}[t]
    \centering
	\subfloat[Splash2]{
    \includegraphics[width=\columnwidth]{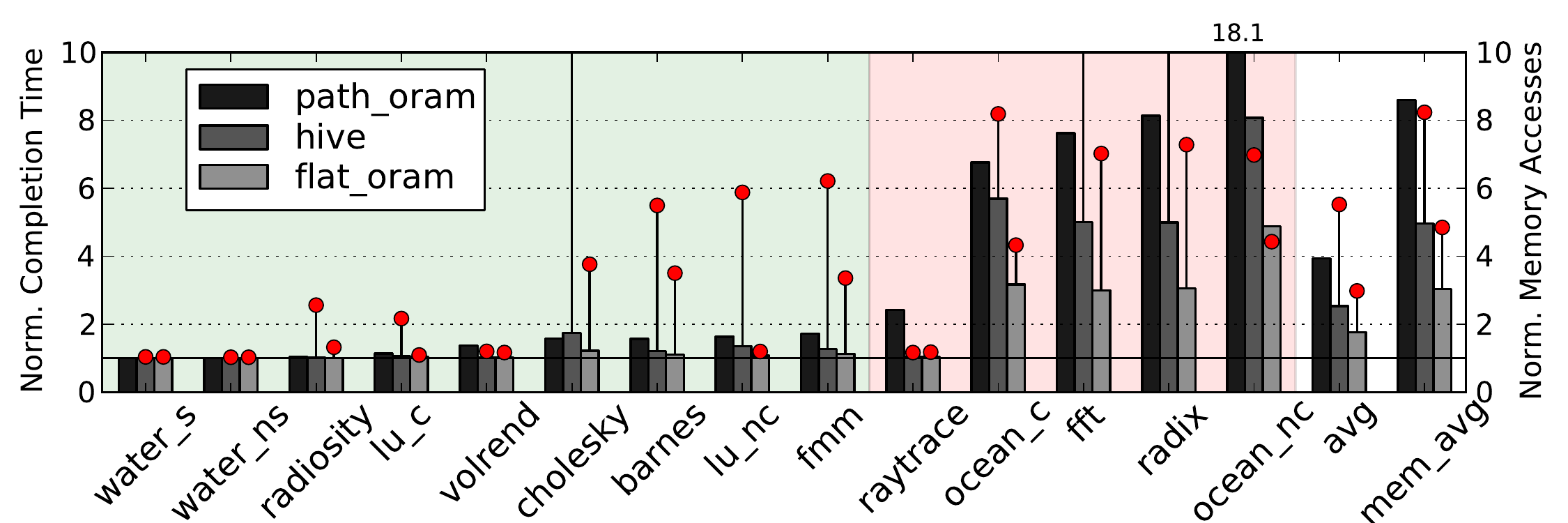}
		\label{fig:main_splash}
	} \\ 
	\vspace{-6pt}
	\subfloat[SPEC06]{
    	\includegraphics[width=\columnwidth]{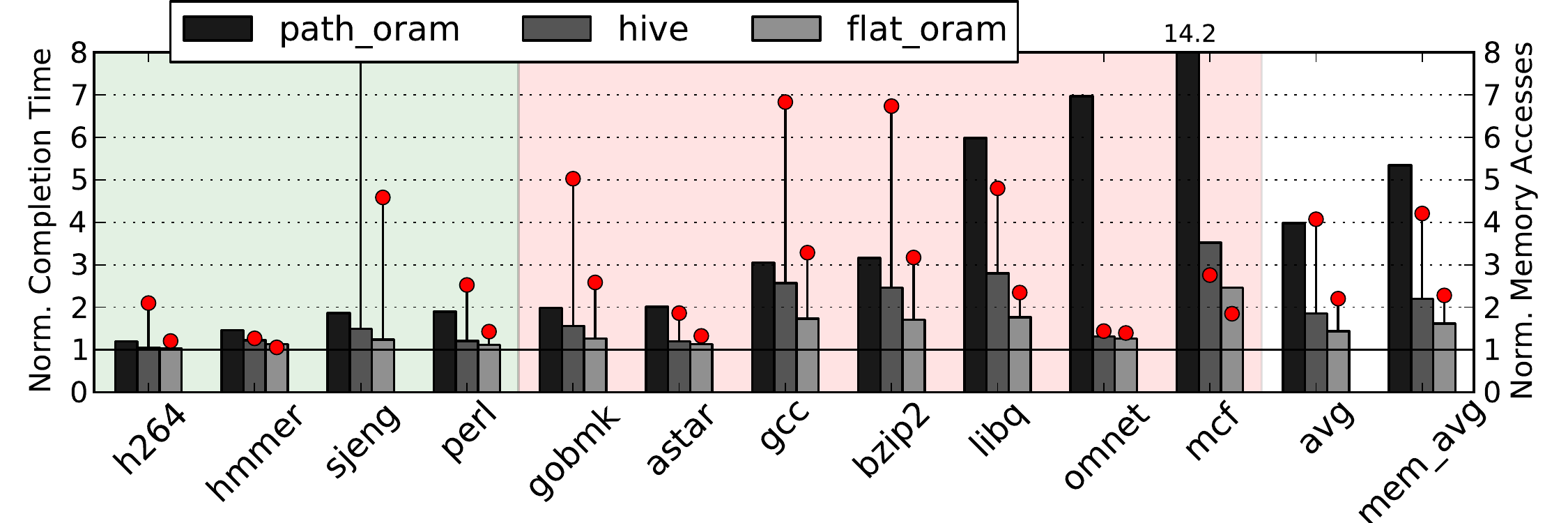}
		\label{fig:main_spec}
	} \\ 
	\vspace{-6pt}
	\subfloat[DBMS]{
		\includegraphics[width=2in]{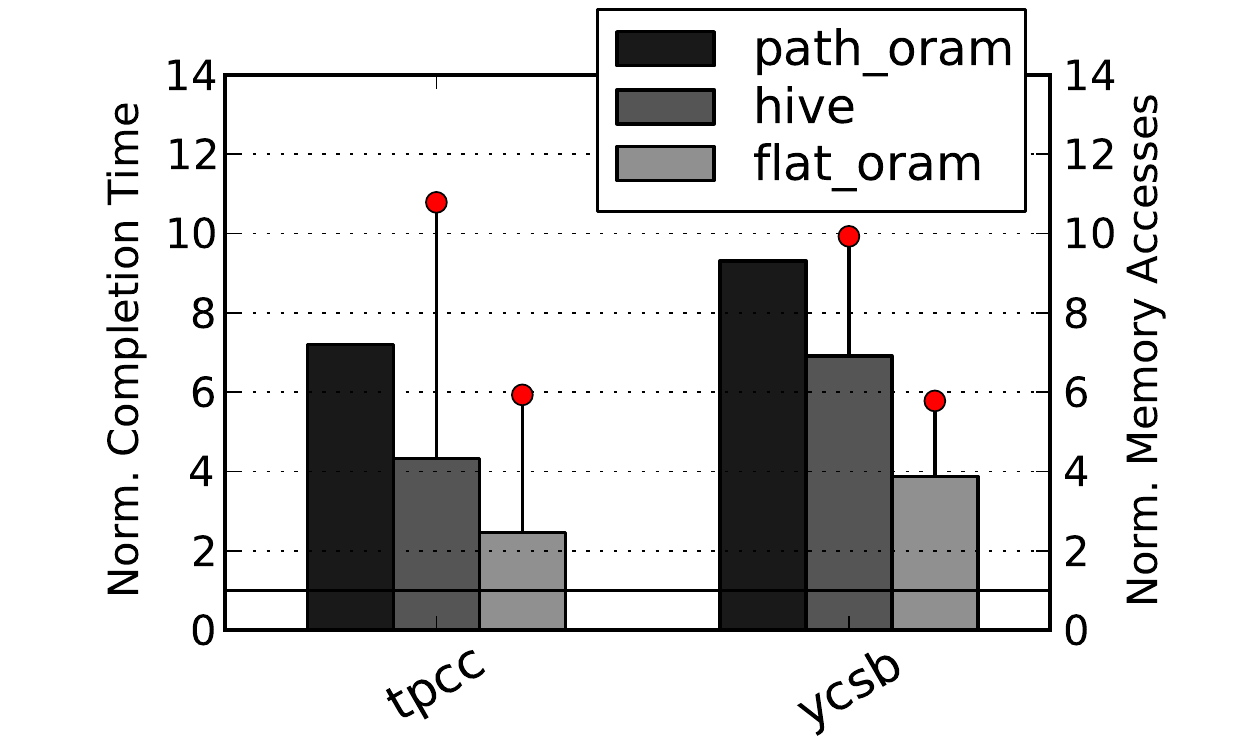}
		\label{fig:main_dbms}
	}
	\vspace{-6pt}
	\caption{Normalized Completion Time and Memory Accesses with respect to Insecure DRAM.}
	\label{graph:mainexp}
	\vspace{-8pt}
\end{figure}

\begin{figure}[t]
    \centering
	\subfloat[Splash2]{
    \includegraphics[width=\columnwidth]{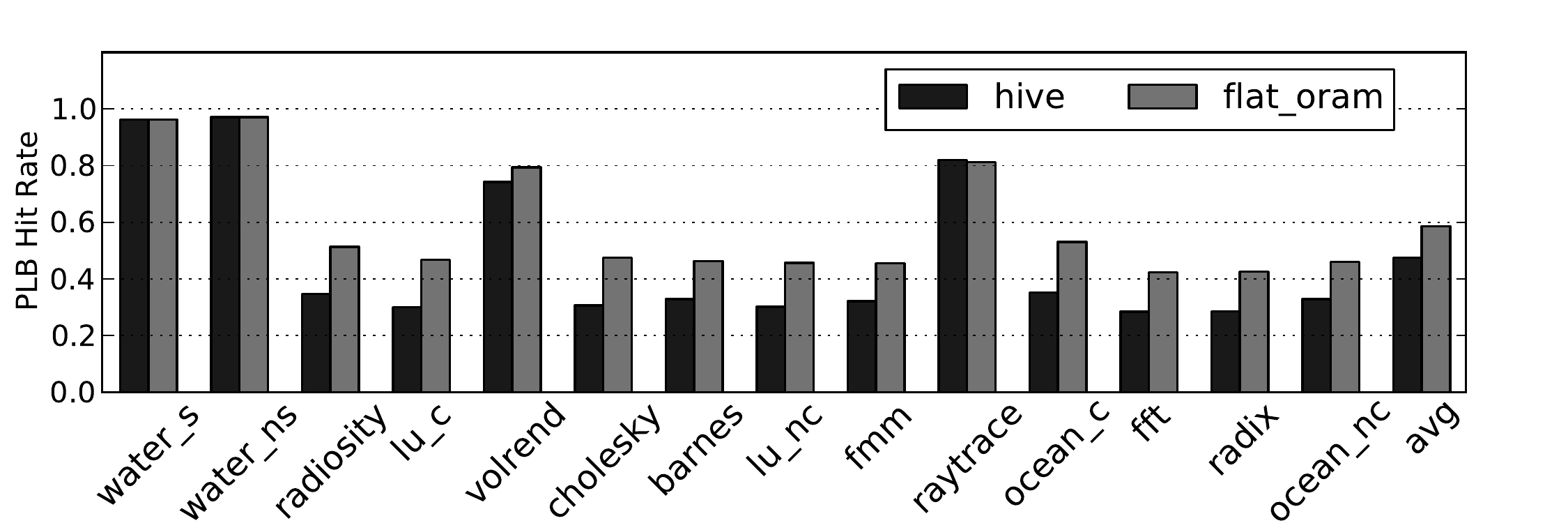}
		\label{fig:plb_splash}
	} \\  
	\vspace{-6pt}
	\subfloat[SPEC06]{
    	\includegraphics[width=\columnwidth]{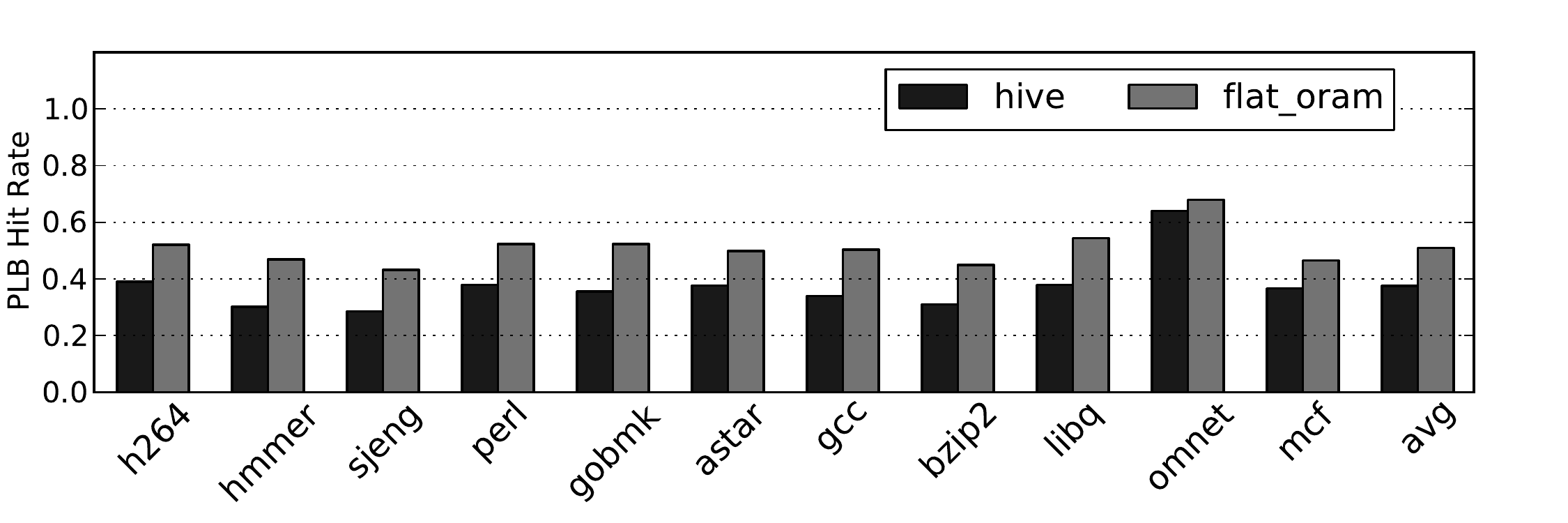}
		\label{fig:plb_spec}
	} \\ 
	\vspace{-6pt}
	\caption{Overall PLB Hit Rate (PosMap blocks and OccMap blocks).}
	\label{graph:plbhit}
	\vspace{-8pt}
\end{figure}

Although all ORAM schemes incur performance slowdown over DRAM, however it is important to note that this slowdown is proportional to the memory intensiveness of the application. 
Memory bound applications suffer from higher performance degradation than compute bound applications.
\figurename~\ref{fig:main_splash}, \figurename~\ref{fig:main_spec} and \figurename~\ref{fig:main_dbms} show normalized completion times (shown by solid bars) of Splash2, SPEC06 and DBMS benchmarks with respect to DRAM.
Splash2 and SPEC06 benchmarks are sorted in ascending order of slowdowns over DRAM from left to right.
We consider all the benchmarks with less than 2$\times$ overhead as \textit{Computation Intensive} benchmarks (plotted over green background) and all those with more than 2$\times$ overhead as \textit{Memory Intensive} benchmarks (plotted over red background).

Clearly, Path ORAM incurs the highest overhead, as expected, among all three ORAM schemes because it is a fully functional ORAM which provides higher security.
However, the point of presenting this comparison is to convince the readers that using Path ORAM for only write-access protection is indeed an overkill when better alternatives (e.g. HIVE, Flat ORAM) exist.
On average, Path ORAM incurs about $8.6\times$ slowdown for Splash2 and $5.3\times$ for SPEC06 memory intensive benchmarks (\textsf{mem\_avg}).
TPCC and YCSB incur $7.2\times$ and $9.3\times$ slowdowns respectively.

HIVE also shows significant performance degradation compared to Flat ORAM for memory intensive benchmarks (\textsf{ocean\_contiguous, ocean\_non\_contiguous, mcf, tpcc, ycsb}).
The average slowdown of HIVE adaptation for memory bound Splash2 and SPEC06 workloads even after several additional optimizations is $5\times$ and $2.2\times$ respectively.
Whereas Flat ORAM outperforms HIVE by up to $50\%$ performance gain on average, having respective average slowdowns of $2\times$ and $1.6\times$.
For DBMS, the performance gain of Flat ORAM over HIVE approach up to $75\%$.

This performance gap is primarily because the inverse position map approach of HIVE results in significantly increased number of additional DRAM accesses.
\figurename~\ref{graph:mainexp} also shows normalized total number of DRAM accesses w.r.t. insecure DRAM system (shown by red markers) for HIVE and Flat ORAM.
These numbers include both the DRAM accesses issued to serve regular ORAM requests and also the ones caused by background evictions (cf. Section~\ref{sec:background-evict}).
The normalized access count for Path ORAM is around $200$ on average, and is not shown on the plots.
It can be seen that HIVE issues $8.2$ and $4.2$ DRAM accesses as opposed to Flat ORAM's $4.8$ and $2.3$ accesses on average for each request issued by the processor for memory intensive Splash2 and SPEC06 workloads respectively.

The reason for higher number of DRAM accesses from HIVE can be found in \figurename~\ref{graph:plbhit} which shows the overall PLB hit rate of both HIVE and Flat ORAM.
The large memory footprint of the HIVE's inverse position map structure results in overall more data being inserted into the PLB and hence translates into higher number of evictions from PLB.
Consequently the ORAM controller experiences higher number of PLB misses and issues relatively higher number of DRAM accesses.
Whereas the dense structure of OccMap offers a smaller memory footprint, thus causing less PLB evictions and exhibiting better locality (cf. Section~\ref{sec:occmap_locality}) which directly translates into performance gain.

The normalized number of DRAM accesses is proportional to the energy consumption of the memory subsystem.
I.e., a higher number of DRAM accesses would result in more energy consumption.
On average, Flat ORAM saves up to $80\%$ energy over HIVE for various workloads.

\subsection{Sensitivity Study} \label{sec:sensitivity}

In this section, we will study how different parameters in the system 
affect the performance of write-only ORAMs.

\paragraph{DRAM Utilization:}
\begin{figure}[t]
	\centering
	\subfloat[ycsb]{
		\includegraphics[height=1.2in]{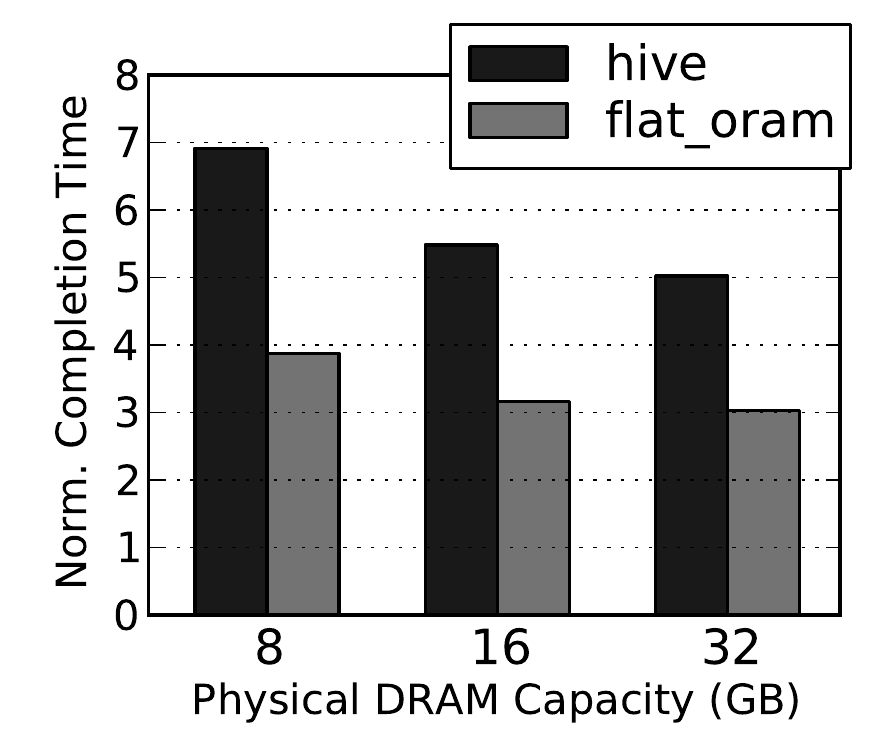}
		\label{fig:oramsize_ycsb}
	}
	\hfil
	\subfloat[sjeng]{
		\includegraphics[height=1.2in]{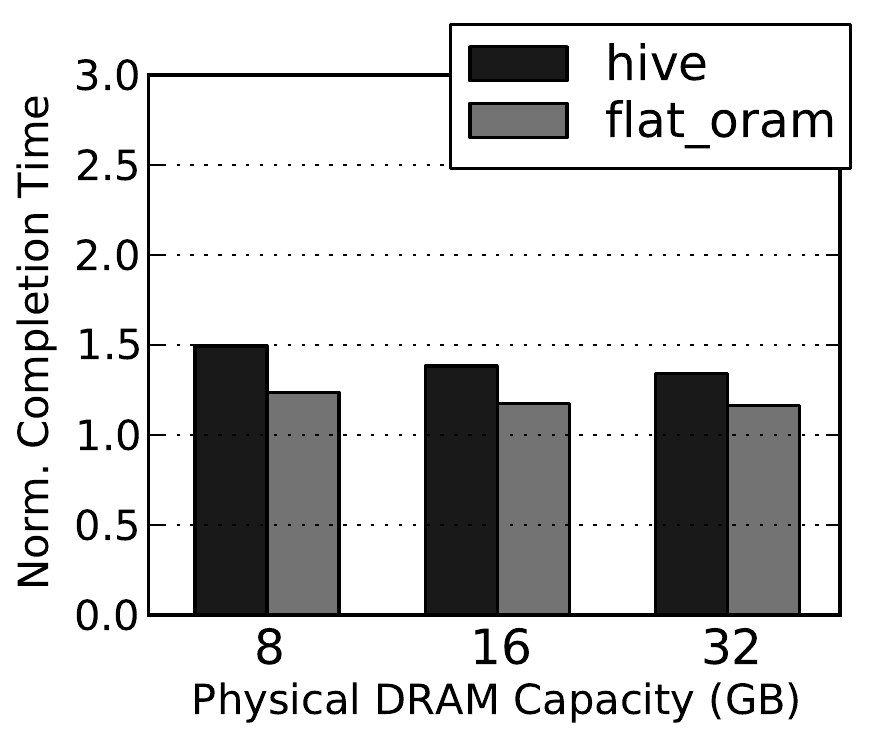}
		\label{fig:oramsize_sjeng}
	}
	\caption{Sweep Physical DRAM Capacity. }
	\label{fig:sweep_oramszie}
	\vspace{-8pt}
\end{figure}
When a block needs to be written to the DRAM, a random position is chosen and if that location is vacant, the block is written at that location (cf. Section~\ref{sec:basic-algo}).
The probability that a randomly chosen location is `vacant' is determined by the DRAM utilization, i.e. the ratio of occupied blocks to total blocks in DRAM.
In order to study the effect of DRAM utilization, we show the results of various physical DRAM sizes ($8$, $16$, $32GB$) for a constant working set of $4GB$ in \figurename~\ref{fig:sweep_oramszie}.
The resulting DRAM utilizations are $50\%$, $25\%$ and $12.5\%$ respectively.

Going from $50\%$ to $25\%$ utilization, memory intensive benchmarks  (\textsf{ycsb}) gain performance, as the collisions during write operations are reduced by half.
However, the jump from $25\%$ to $12.5\%$ utilization yields little gain because the collision probability of $25\%$ at $16GB$ mark is already too low to be a major performance bottleneck.
Notice that HIVE benefits more compared to Flat ORAM from the reduced collisions since it has a much higher collision-penalty.
Since less memory intensive benchmarks (\textsf{sjeng}) are not constrained by write operations anyway, lower utilizations do not help much.

\paragraph{Stash Size:}
\begin{figure}[t]
	\centering
	\subfloat[ocean\_non\_contiguous]{
		\includegraphics[height=1.2in]{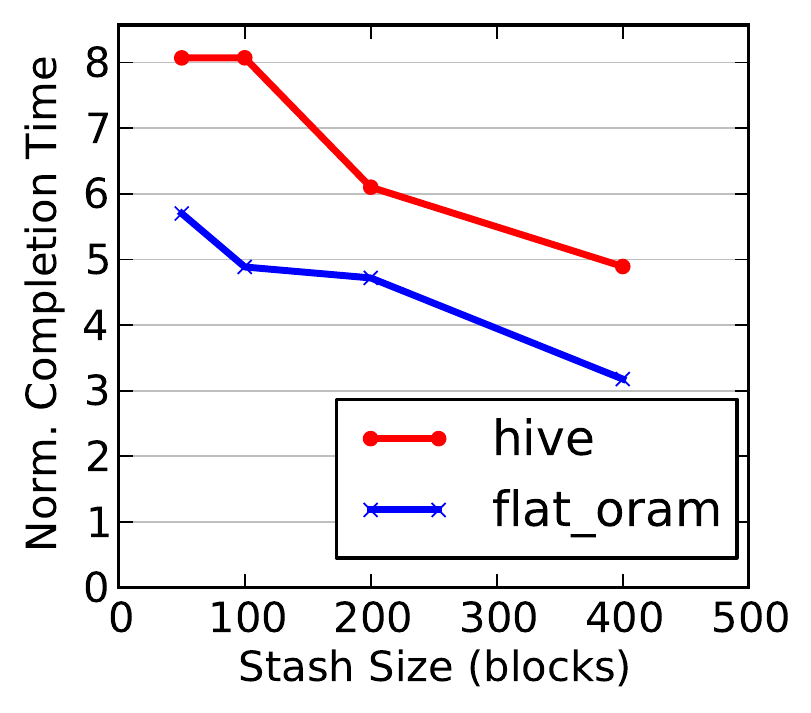}
		\label{fig:stash_ocean_non_c}
	}
	\hfil
	\subfloat[sjeng]{
		\includegraphics[height=1.2in]{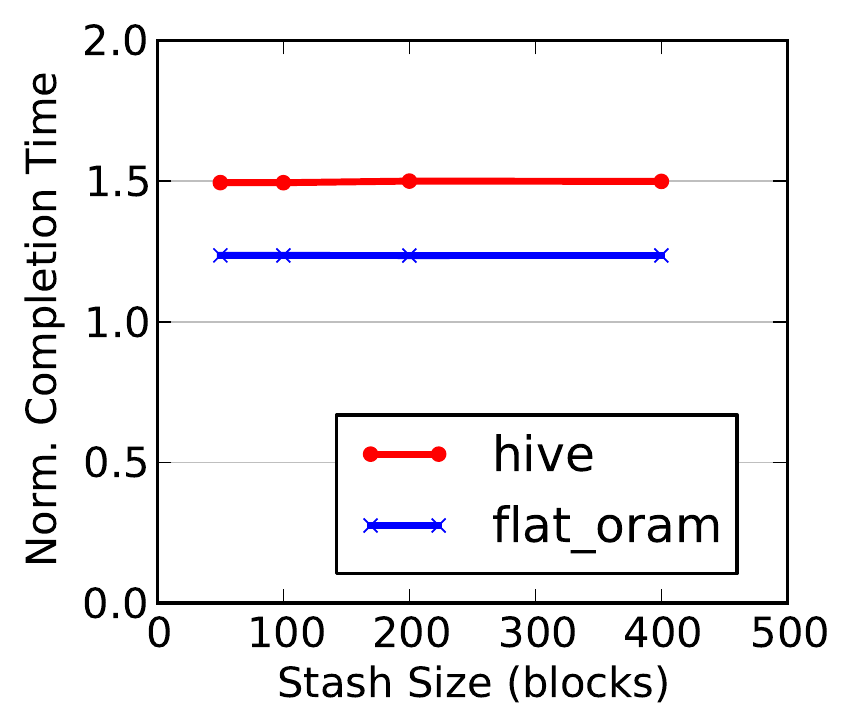}
		\label{fig:stash_sjeng}
	}
	\caption{ Sweep Stash Size. }
	\label{fig:sweep_stashsize}
	\vspace{-8pt}
\end{figure}
As discussed in Section~\ref{sec:background-evict}, when the stash occupancy increases than a particular threshold, the ORAM starts performing `background evictions'.
Since background evictions cause the real requests to be suspended temporarily, frequent background evictions cause performance degradation.
A larger stash is less likely to become full and thus reduces background eviction rate and improves performance. 

In Figure~\ref{fig:sweep_stashsize}, the stash size is swept for two different benchmarks, one is highly memory intensive (\textsf{ocean\_non\_contiguous}) and the other one is significantly less memory bound (\textsf{sjeng}).  
The memory intensive benchmark benefits from a large stash, as it experiences high background evictions rate at lower stash sizes.
The less memory intensive benchmark does not benefit much from increased stash sizes, as it already has a low background evictions rate.
In general, Flat ORAM shows significant performance gain over HIVE even at small stash sizes.


\paragraph{DRAM Latency \& Bandwidth:}
\begin{figure}[t]
	\centering
	\subfloat[ocean\_contiguous]{
		\includegraphics[height=1.2in]{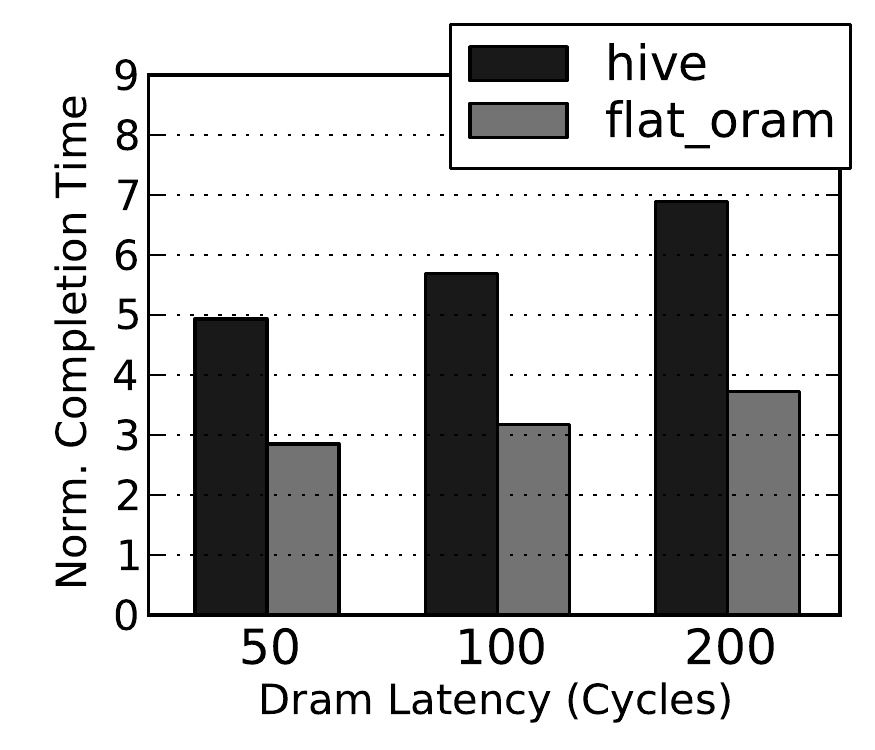}
		\label{fig:latency_ocean_c}
	}
	\hfil
	\subfloat[sjeng]{
		\includegraphics[height=1.2in]{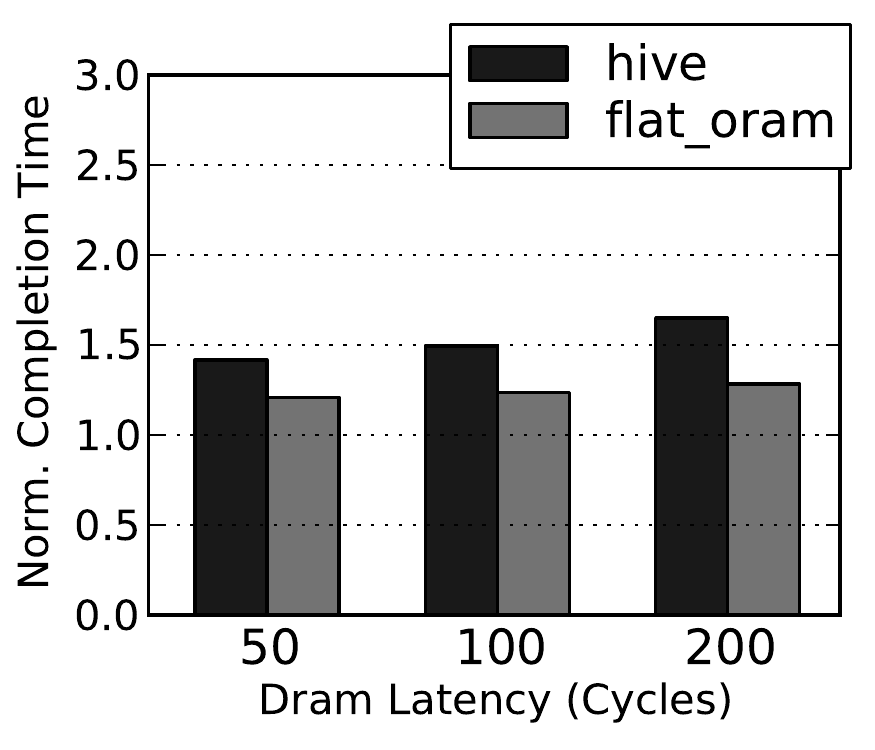}
		\label{fig:latency_sjeng}
	}
	\caption{ Sweep DRAM latency. }
	\label{fig:sweep_latency}
	\vspace{-8pt}
\end{figure}

\begin{figure}[t]
	\centering
	\subfloat[ocean\_contiguous]{
		\includegraphics[height=1.2in]{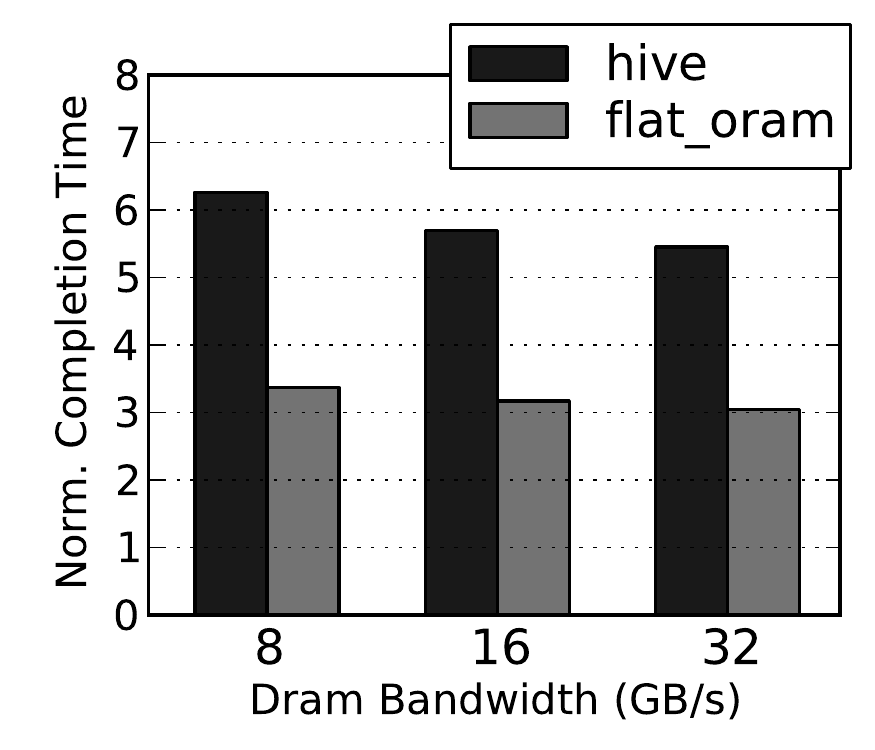}
		\label{fig:band_ocean_c}
	}
	\hfil
	\subfloat[sjeng]{
		\includegraphics[height=1.2in]{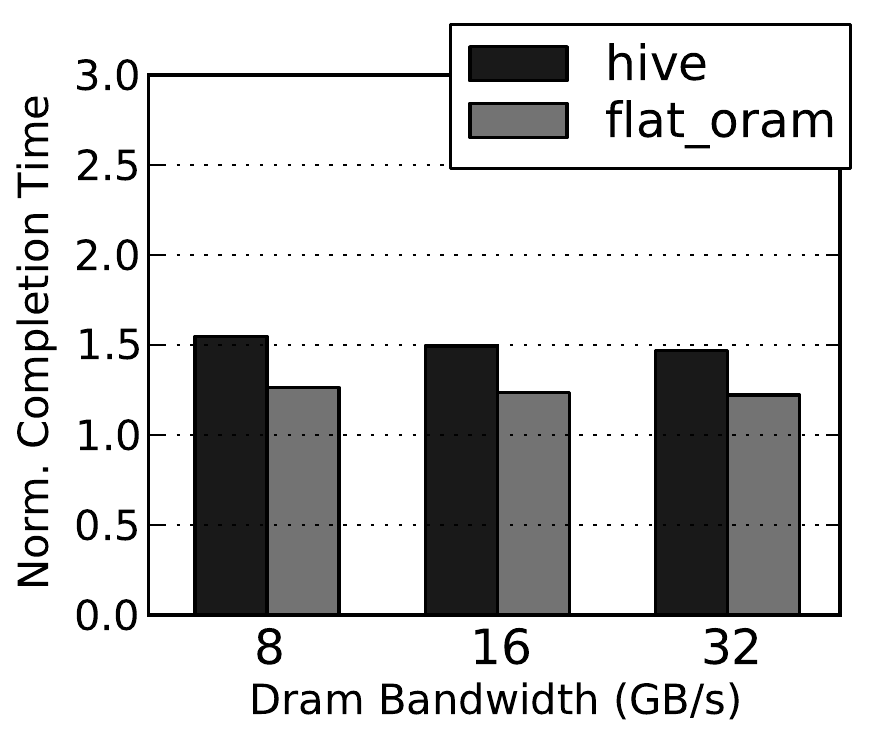}
		\label{fig:band_sjeng}
	}
	\caption{ Sweep DRAM bandwidth. }
	\label{fig:sweep_band}
	\vspace{-8pt}
\end{figure}
In Path ORAM, each ORAM access results in about $200$ DRAM accesses on average under typical parameter settings.
Most of these accesses can be issued in a burst without waiting for the first data block to arrive, since the addresses are known a priori, e.g. accessing a full path.
Therefore, the DRAM bandwidth becomes the main bottleneck in Path ORAM, whereas the DRAM latency plays less significant role as it is incurred less often.

However, the write-only ORAMs under consideration typically only issue less than $10$ DRAM accesses per ORAM access (cf. Section~\ref{sec:eval-splash2}).
Furthermore, there could be interdependencies within these $10$ accesses, e.g., reading an OccMap block to find out if a position is vacant, and then issuing further writes in case a vacant position is found.
In such cases, DRAM latency is incurred more often and hence plays more prominent role in the overall performance than the DRAM bandwidth.

This phenomenon is shown in DRAM latency and bandwidth sweep studies in \figurename~\ref{fig:sweep_latency} and \figurename~\ref{fig:sweep_band} respectively.
Memory intensive benchmarks (\textsf{ocean\_contiguous}) are more sensitive to DRAM latency and experience more performance degradation at higher latencies.
On the other hand, compute bound benchmarks (\textsf{sjeng}) are less sensitive to the DRAM latency.
Increasing the DRAM bandwidth seems to help only a little as expected and explained in the discussion above.

\vspace{-8pt}
\paragraph{Periodic ORAM:} \label{sec:eval-periodicity}

\begin{figure}[t]
	\centering
	\subfloat[Splash2]{
		\includegraphics[width=\columnwidth]{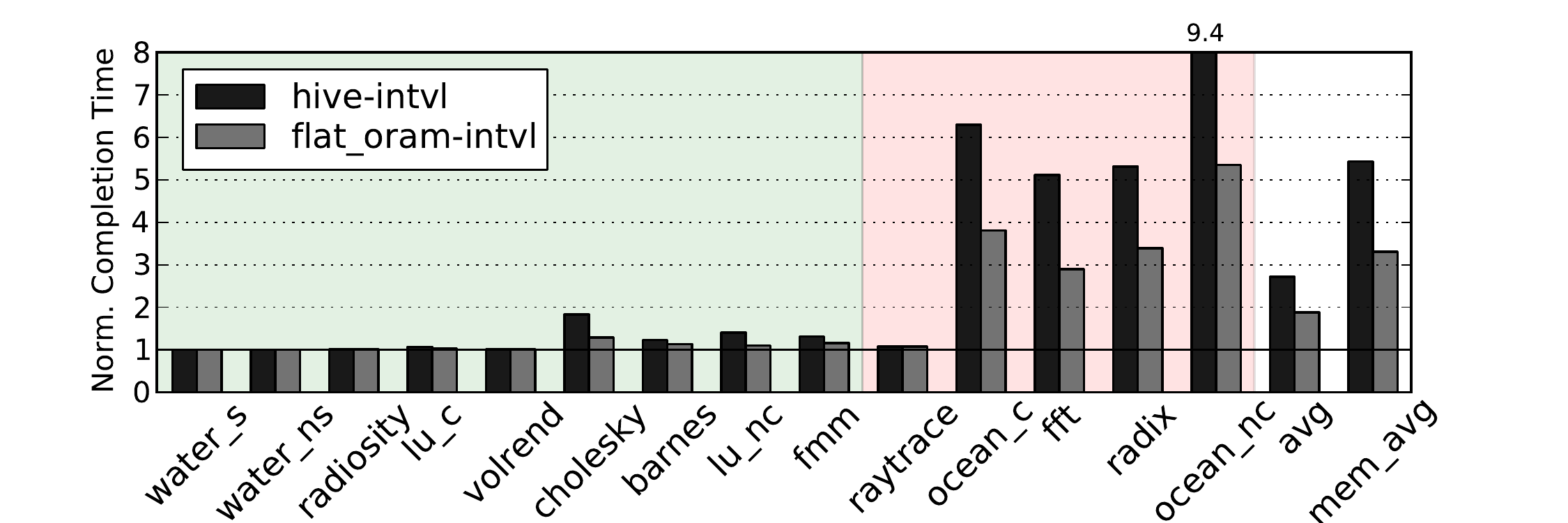} } \\ 
	\vspace{-6pt}
	\subfloat[SPEC06]{
		\includegraphics[width=\columnwidth]{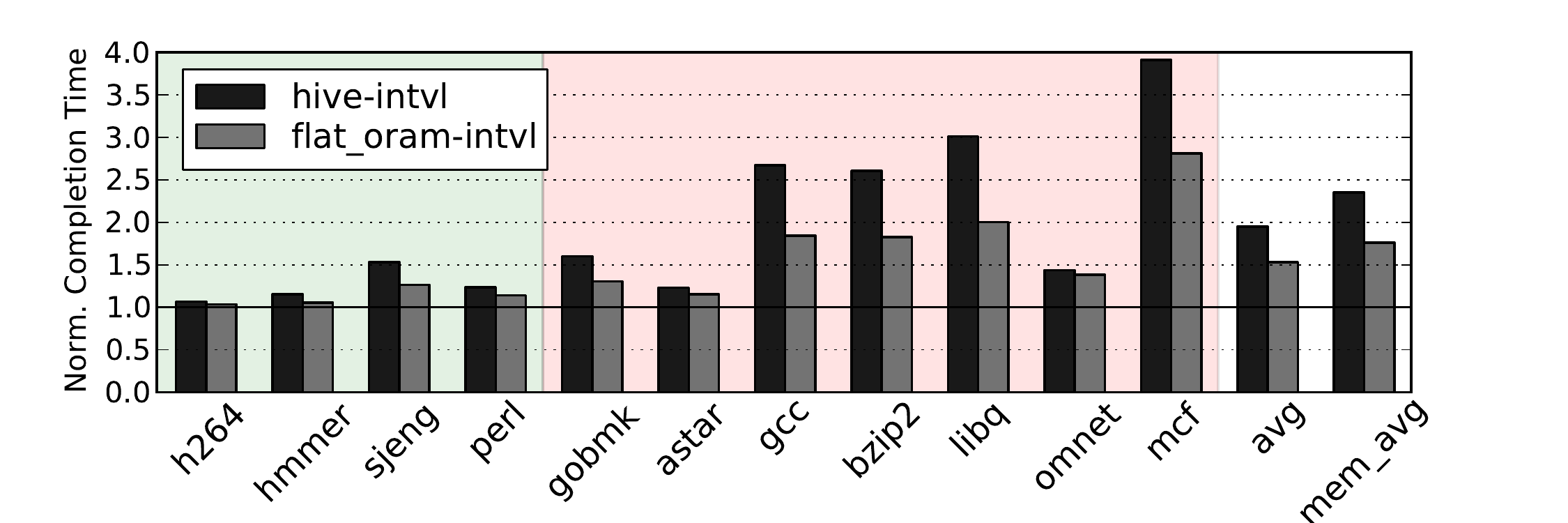} } \\ 
	\caption{Periodic ORAM accesses. Normalized w.r.t. Insecure DRAM. ORAM Period = 100 cycles.}
	\label{fig:periodic}
	\vspace{-6pt}
\end{figure}


Figure~\ref{fig:periodic} shows the experimental results of periodic write-only ORAM schemes. 
The results are normalized to insecure DRAM.
The period in terms of number of cycles between two consecutive ORAM accesses is chosen to be 100 cycles.
In general, adding periodicity to our ORAM scheme does not significantly hurt performance.

\section{Conclusion} \label{sec:conclusion}

We propose an efficient and practical write-only Oblivious RAM scheme called \textit{Flat ORAM} for secure processor architectures.
It is the first write-only ORAM with a concrete implementation in secure processors domain.
The implementation details are discussed and the design space is comprehensively explored.  
On memory intensive Splash-2 and SPEC06 benchmarks, Flat ORAM only incurs on average $3\times$ and $1.6\times$ slowdown respectively.
Compared to a closest related work in the literature, Flat ORAM offers up to $75\%$ higher performance and $80\%$ energy savings.

\begin{acks}
The work is partially supported by NSF grant CNS-1413996 for MACS: A Modular Approach to Cloud Security.
\end{acks}

\bibliographystyle{ACM-Reference-Format}
\bibliography{sections/refs,sections/main,sections/generic,sections/security,sections/onChipNetwork,sections/local,sections/hw,sections/bib-sharon}


\end{document}